\documentstyle[12pt]{article}

\newcommand{\ds}{\displaystyle}
\newcommand{\ol}{\overline}
\newcommand{\be}{\begin{equation}}
\newcommand{\ee}{\end{equation}}
\newcommand{\ba}{\begin{eqnarray}}
\newcommand{\ea}{\end{eqnarray}} 
\newcommand{\lb}[1]{\label{#1}}  
\newcommand{\bb}[1]{\bibitem{#1}}

\newcommand{\Sch}{Schwarzschild }
\begin{document}
\begin{titlepage}
\setcounter{page}{1}
\title{
\hfill {\small DF--UFES 002/97}\\
\vskip15mm
From Schwarzschild to Kerr: \\ 
Generating spinning Einstein--Maxwell fields \\ from static fields
}
\author{
\bigskip
G\'erard Cl\'ement\thanks{Permanent address: 
Laboratoire de Gravitation et Cosmologie Relativistes,
Universit\'e Pierre et Marie Curie, CNRS/URA769,
Tour 22-12, Bo\^{\i}te 142, 4 place Jussieu,
75252 Paris cedex 05, France. 
E-mail: gecl@ccr.jussieu.fr.}\\
\small Departamento de F\'{\i}sica,\\
\small Universidade Federal do Espir\'{\i}to Santo,\\
\small Vit\'oria, ES, Brazil
}
\bigskip
\date{October 23, 1997}
\maketitle

\begin{abstract}
The Kerr solution is generated from the \Sch solution by a simple combination of real global coordinate transformations and of invariance transformations acting on the space of stationary solutions of the Einstein--Maxwell equations. The same transformation can be used to generate a spinning field configuration from any static axisymmetric 
configuration. We illustrate this by generating from the continuous family of Voorhees--Zipoy vacuum solutions a family of solutions endowed with mass, angular momentum, dipole magnetic moment and quadrupole electric moment.
\end{abstract}

\end{titlepage}

The four--dimensional stationary Einstein--Maxwell equations are well--known to be invariant under an SU(2,1) group of transformations \cite{nk}\cite{ex}. Applied to asymptotically flat solutions, these transformations map monopole solutions into monopole solutions (e.g. the \Sch solution into the Reissner--Nordstr\"{o}m solution), so that the invariance group SU(2,1) cannot be used to generate a monopole--dipole solution, such as the Kerr solution, from a (static) monopole solution, such as the \Sch solution. 

The situation becomes different in the case of stationary axisymmetric solutions ---solutions to the Einstein--Maxwell equations with two commuting Killing vectors. By combining the invariance transformations associated with an arbitrarily chosen direction in the two--Killing vector space with transitions to other linear combinations of the two Killing vectors, the infinite--dimensional Geroch group emerges \cite{geroch}. The infinitesimal action of this group was exponentiated by Kinnersley and Chitre \cite{kc1}, who also identified an infinite--parameter subgroup which preserves asymptotic flatness \cite{kc2} and which, for instance, may be used to generate the Kerr solution from the \Sch solution. These transformations allow in principle the generation of all solutions of the stationary axisymmetric Einstein--Maxwell problem, which is thus completely integrable. This generation of stationary axisymmetric solutions can also be achieved through standard inverse--scattering !
transform methods applied to the E
instein--Maxwell problem \cite{maison}. 

Curiously, the direct generation of rotating solutions from static axisymmetric solutions by combining finite SU(2,1) transformations with finite coordinate transformations mixing the two Killing vectors has not to our knowledge been attempted up to now, probably because such coordinate transformations do not preserve asymptotic flatness. In this paper, we shall show that an appropriate combination of SU(2,1) and coordinate transformations may be used for the direct generation of asymptotically flat spinning Einstein--Maxwell fields from asymptotically flat static fields. First, we shall generate the Kerr solution from the \Sch 
solution by combining a real coordinate transformation to a frame rotating at uniform velocity ---which by itself does not lead to a new solution--- with specific SU(2,1) transformations. Then, we shall generalize our construction to the case of an arbitrary input static axisymmetric field, and apply it to the generation of a continuous family of spinning solutions from the Voorhees--Zipoy \cite{zip}\cite{voor} family of static solutions. 

We first review briefly the potential space approach to the stationary
Einstein--Maxwell equations \cite{ex}.  The
metric for spacetimes admitting a timelike Killing vector may be
parametrized by
\be\lb{stat1}
ds^2 = f\,(dt - \omega_i dx^i)^2 - f^{-1}\,h_{ij}\,dx^i dx^j\,
\ee
where the fields $f$, $\omega_i$ and $h_{ij}$ depend only on the space
coordinates $x^i$. The electromagnetic field $F_{\mu\nu}$ may be
parametrized by two electric and magnetic scalar potentials $v$ and $u$ 
such that
\be\lb{stat2}
F_{i0} = \partial_i v\,, \qquad F^{ij} = f\,h^{-1/2}\epsilon^{ijk}
\partial_k u\,.
\end{equation}
Finally, the twist or gravimagnetic potential $\chi$ is defined
\be\lb{twist}
\partial_i\chi = -f^2\,h^{-1/2}h_{ij}\,\epsilon^{jkl}\partial_k\omega_l 
+ 2(u\partial_i v - v\partial_i u)\,.
\end{equation}
The complex Ernst potentials \cite{er} are related to the four real scalar
potentials $f$, $\chi$, $v$ and $u$ by
\be
{\cal E} = f + i \chi - \ol{\psi}\psi\,, \qquad \psi = v + iu\,.
\ee
The stationary Einstein--Maxwell equations then reduce to the
three--dimensional Ernst equations \cite{er}
\ba\lb{ernst2}
f\nabla^2{\cal E} & = & \nabla{\cal E} \cdot (\nabla{\cal E} + 
2\ol{\psi}\nabla\psi)\,,\nonumber \\
f\nabla^2\psi & = & \nabla\psi \cdot (\nabla{\cal E} + 
2\ol{\psi}\nabla\psi)\,, \\
f^2R_{ij}(h) & = & {\em Re} 
\left[ \frac{1}{2}{\cal E},_{(i}\ol{{\cal E}},_{j)} 
+ 2\psi{\cal E},_{(i}\ol{\psi},_{j)}
-2{\cal E}\psi,_{(i}\ol{\psi},_{j)} \right]\,, \nonumber
\ea
where the scalar products and Laplacian are computed with the metric $h_{ij}$.

These equations are invariant under an SU(2,1) group of transformations
\cite{nk}. The Ernst potentials may be expressed in terms of three new complex 
Kinnersley potentials $U$, $V$, $W$ (one of which is redundant) by \cite{kin}
\be\lb{vect}
{\cal E} = \frac{U-W}{U+W}\,, \qquad \psi = \frac{V}{U+W}\,.
\ee
The SU(2,1) transformations acting linearly on the complex vector $(U,V,W)$
and leaving invariant the norm $\ol{U} U + \ol{V} V - \ol{W} W$
transform solutions of equations (\ref{ernst2}) with the spatial metric 
$h_{ij}$ into other solutions with the same spatial metric. We shall use
in the following the SU(2,1) involution 
\be\lb{inv}
\Pi: \quad U \leftrightarrow V\,,
\ee
which transforms a solution $({\cal E}, \psi)$ into a solution $(\hat{\cal
E}, \hat{\psi})$ with
\be\lb{inv2}
\hat{\cal E} = \frac{-1 + {\cal E} + 2 \psi}{1 - {\cal E} + 2 \psi}\,, \qquad 
\hat{\psi} = \frac{1 + {\cal E}}{1 - {\cal E} + 2 \psi}\,,
\ee
and a vacuum solution ($\psi = 0$) into a solution with $\hat{\cal E} = -1$.
 
We choose as our starting point the \Sch solution
\be
ds^2 = (1-\frac{2m}{r})\,dt^2 - (1-\frac{2m}{r})^{-1}\,dr^2 -r^2(d\theta^2
+ \sin^2\theta\,d\varphi^2)\,,
\ee
which may be rewritten in the form (\ref{stat1}) as
\be\lb{S}
ds^2 = fdt^2 - f^{-1}m^2\,[dx^2 + (x^2-1)(d\theta^2 + 
\sin^2\theta\,d\varphi^2)]\,, \qquad f = \frac{x-1}{x+1}\,,
\ee
with $x = (r-m)/m$.
The associated Ernst potentials may be put in the form (\ref{vect}) with
$U=x$, $V=0$, $W= 1$. Now let us act on the \Sch solution with the SU(2,1)
transformation (\ref{inv}). The resulting static 
solution $\hat{\psi} = x$, $\hat{f} = x^2-1$ is the open Bertotti--Robinson 
solution \cite{br} 
\be\lb{BR}
d\hat{s}^2 = m^2 \left[ (x^2-1)\,d\tau^2 - \frac{dx^2}{x^2-1}  - 
\frac{dy^2}{1-y^2} - (1-y^2)\,d\varphi^2 \right],
\ee
with $\tau = m^{-1}t$, $y = \cos\theta$. This non-asymptotically flat 
spacetime can be viewed as resulting from spontaneous
compactification of four--dimensional spacetime to the direct product 
of two constant curvature two--dimensional spaces, an ``external'' 
anti--de Sitter space adS$_2$, and an internal spherical space S$^2$.

Our second step is to transform the static Bertotti--Robinson solution to
a uniformly rotating frame. Starting from a generic stationary
axisymmetric metric
\be
ds^2 = f(\,dt - \omega\,d\varphi)^2 - f^{-1}\,(\gamma_{mn}\,dx^m\,dx^n +
\rho^2\,d\varphi^2) 
\ee 
(where $m,n = 1,2$, and ${\rho,_m}^{;m} = 0$), such a global coordinate 
transformation
\be\lb{rot1}
d\varphi = d\varphi' + \Omega\,dt
\ee
leads to the new metric functions
\ba\lb{rot2}
f' & = & f\,[1 - 2\Omega\omega + \Omega^2(\omega^2 - f^{-2}\rho^2)]\,,
\nonumber \\
\omega' & = & \frac{\omega - \Omega(\omega^2 - f^{-2}\rho^2)}
{1 - 2\Omega\omega + \Omega^2(\omega^2 - f^{-2}\rho^2)}\,, \\
f'^{-1}\gamma'_{mn} & = & f^{-1}\gamma_{mn}\,, \qquad \rho' = \rho\,. \nonumber 
\ea
A simple consequence of these relations is
\be
f'(1+\Omega\omega') = f(1-\Omega\omega)\,.
\ee
The corresponding transformed electromagnetic field components are
\ba
F'_{m0} & = & F_{m0} + \Omega\,F_{m3}\,, \nonumber \\
F'_{m3} & = & F_{m3}\,,
\ea
from which, using (\ref{stat2}), we obtain the relations giving the new
scalar potentials $v'$ and $u'$
\ba\lb{rot3}
\partial_m v' & = &  (1-\Omega\omega)\partial_m v - \Omega f^{-1} \rho \,
\tilde{\partial}_m u \,, \nonumber \\
\partial_m u' & = &  (1-\Omega\omega)\partial_m u + \Omega f^{-1} \rho \,
\tilde{\partial}_m v \,. 
\ea 
with 
$$\tilde{\partial}_m = \gamma^{-1/2} \gamma_{mn} \epsilon^{np} \partial_p\,.$$
In the case of an electrostatic metric with $\omega = 0$, $u = 0$, $v =
\psi$, the relations (\ref{rot2}) and (\ref{rot3}) simplify to
\ba
f' = f - \Omega^2 \frac{\rho^2}{f} & , & \qquad \omega' = \Omega
\frac{\rho^2}{f f'}\,, \nonumber \\
\partial_m v' = \partial_m \psi & , & \partial_m u' = \Omega \frac{\rho}{f}  
\tilde{\partial}_m \psi \,,
\ea
from which we obtain the relations giving the new Ernst potentials  
\ba
\partial_m \psi' & = & D_m \psi \,, \nonumber \\
\partial_m {\cal E}' & = & D_m {\cal E} + i \left[ -\Omega \frac{f}{\rho}
\tilde{D}_m(\frac{\rho^2}{f}) + 2u' D_m \psi \right] ,
\ea
with
$$D_m \equiv \partial_m + i \Omega  \frac{\rho}{f} \tilde{\partial}_m \,.$$

Applying this transformation to the Bertotti--Robinson solution
(\ref{BR}), which is such that
\be
\rho^2 = m^2 (x^2-1) (1-y^2)\,,
\ee
we obtain the new metric functions
\ba
\hat{f}' & = & x^2 + m^2 \Omega^2 y^2 - (1+m^2\Omega^2) \,, \nonumber \\
\hat{\omega}' & = & \frac{m^2 \Omega (1-y^2)}{x^2 + m^2 \Omega^2 y^2 -
(1+m^2\Omega^2)} \,, 
\ea
and Ernst potentials
\be
\hat{\cal E}' = -(1+m^2\Omega^2)\,, \qquad \hat{\psi}' = x - im\Omega y\,.
\ee
This is almost in the class ${\cal E} = -1$ of solutions, and can be
transformed to this class by a time dilation
\be\lb{dil}
t' = (1+m^2\Omega^2)^{-1/2}t''\,,
\ee
leading to $\hat{\cal E}'' = (1+m^2\Omega^2)^{-1}\hat{\cal E}'$, 
$\hat{\psi}'' = (1+m^2\Omega^2)^{-1/2}\hat{\psi}'$, i.e.
\be\lb{BR2} 
\hat{\cal E}'' = -1\,,\qquad \hat{\psi}'' = px-iqy \qquad (\,(p^2 + q^2) = 1\,)
\ee
with $p = (1+m^2\Omega^2)^{-1/2}$, $q = m\Omega(1+m^2\Omega^2)^{-1/2}$;
note that the coordinate $\rho' = \rho$ is transformed by (\ref{dil}) to 
$\rho'' = p\rho$.

The Ernst potentials (\ref{BR2}) describe again the static Bertotti--Robinson 
spacetime, now 
viewed in a uniformly rotating frame. In a third step, we transform back this
$\hat{\cal E}'' = -1$ solution by the involution (\ref{inv}) to recover a 
vacuum solution
\be\lb{Kerr}
{\cal E} = \frac{px-iqy-1}{px-iqy+1}\,, \qquad \psi = 0\,.
\ee
We recognize in (\ref{Kerr}) the Ernst \cite{er} form of the Kerr solution
with the parameters $m$ and $a = mq$, $x$ and $y$ being the prolate
spheroidal coordinates \cite{zip} related to the Weyl coordinates $\rho$
and $z$ by
\ba
\rho & = & \nu\,(x^2-1)^{1/2} (1-y^2)^{1/2}\,, \nonumber \\
z & = & \nu\,xy\,. 
\ea
with $\nu = mp$.

The charged Reissner--Nordstr\"{o}m solution and its spinning
generalization, the Kerr--Newman solution, are related respectively to the
\Sch solution and the Kerr solution by SU(1,1) boosts ${\cal U}$ in the
$(V, W)$ subspace.
Combining these transformations with the above described transformation
$\Sigma$ which leads from the \Sch solution to the Kerr solution, we are
able to generate the Kerr--Newman solution from the Reissner--Nordstr\"{o}m
solution by the transformation ${\cal U}\,\Sigma\,{\cal U}^{-1}$. 
What is not so obvious
is that the diagram commutes, i.e. that essentially the same operation
$\Sigma$ may be used to transform directly the Reissner--Nordstr\"{o}m
solution into the Kerr-Newman solution. The reason is that the
Bertotti--Robinson solution is essentially unaffected (only rescaled) by
boosts $\hat{\cal U}$ in the $(U, W)$ subspace. The Reissner--Nordstr\"{o}m 
solution
\be
ds^2 = (1-\frac{2m}{r}+\frac{e^2}{r^2})dt^2 - 
(1-\frac{2m}{r}+\frac{e^2}{r^2})^{-1}dr^2 -r^2(d\theta^2 + 
\sin^2\theta\,d\varphi^2)\,, \quad \psi = \frac{e}{r}\,,
\ee
may be rewritten as
\be
ds^2 = fdt^2 - f^{-1}\mu^2\,[dx^2 + (x^2-1)(d\theta^2 + 
\sin^2\theta\,d\varphi^2)]\,, \qquad f = \frac{\mu^2(x^2-1)}{(\mu x+m)^2}\,,
\ee
where $r = \mu x + m$, and $\mu^2 = m^2 - e^2$. The associated Ernst
potentials are
\be
{\cal E} = \frac{\mu x - m}{\mu x + m}\,, \qquad \psi = \frac{e}{\mu x + m}\,.
\ee
The involution (\ref{inv}) transforms these into
\be
\hat{\cal E} = - \lambda^2 \, \qquad \hat{\psi} = \lambda x\,,
\ee
with $ \lambda^2 = (m-e)/(m+e)$. The resulting metric is again a
Bertotti--Robinson metric (\ref{BR}) with $m^2$ replaced by $(m+e)^2$, and
$\tau = \lambda (m+e)^{-1} t$. It follows that the same transformations as
before, a uniform frame rotation combined with an appropriate time
rescaling, result in the ``spinning'' Ernst potentials
\be
\hat{\cal E}'' = - \lambda^2\,, \qquad \hat{\psi}'' = \lambda (px - iqy) \,,
\ee
with $p^2 + q^2 = 1$. The involution (\ref{inv}) then leads to the Ernst
potentials 
\be
{\cal E} = \frac{\mu (px-iqy) - m}{\mu (px-iqy) + m}\,, \qquad \psi =
\frac{e}{\mu (px-iqy) + m}  
\ee
of the Kerr--Newman solution. The time rescaling such that the spinning
Ernst potential $\hat{\cal E}''$ is scaled to the static one ensures that
the monopole parameters of the spinning solution are scaled to the static
values.

Our spin--generating procedure may again be generalized to arbitrary
axisymmetric asymptotically flat static fields. The fields generated by an
arbitrary axisymmetric static configuration of masses and charges are 
asymptotic to 
Reissner--Nordstr\"{o}m fields with parameters $(m, e)$ equal to the
total mass and charge of the system. At each step of our transformations,
the leading asymptotic behavior is governed by these monopole fields, so
that the spinning field configuration generated by the transformation
$\Sigma$ is asymptotic to the original static field configuration. However in the general case this construction has no reason to be invariant under
the SU(1,1) boosts ${\cal U}$. Because the field configuration transformed from the original static field configuration by the involution (\ref{inv}) is only asymptotic to the Bertotti--Robinson configuration, there is no
reason for global coordinate transformations of this configuration to
commute with the boosts $\hat{\cal U}$.

Let us consider in some detail the example of the Voorhees--Zipoy solution
\cite{zip}\cite{voor}. This family of static axisymmetric vacuum metrics
is given in prolate spheroidal coordinates by
\be\lb{VZ}
ds^2 = f\,dt^2 - f^{-1} \nu^2 \left[ {\rm e}^{2k} (x^2-y^2)(\frac{dx^2}{x^2-1}
+ \frac{dy^2}{1-y^2}) + (x^2-1)(1-y^2)\,d\varphi^2 \right] ,
\ee
with
\be
f = {\left( \frac{x-1}{x+1} \right)}^{\delta}\,, \qquad {\rm e}^{2k} =
{\left( \frac{x^2-1}{x^2-y^2} \right)}^{\delta^2}\,, 
\ee
where $\delta$ is a real parameter. The Ernst potentials for this solution 
are given by (\ref{vect}) with the Kinnersley potentials
\be
U = \coth \sigma\,, \quad V = 0\,, \quad W = 1\,,
\ee
where $\sigma = \frac{\delta}{2} \ln \frac{x+1}{x-1}$. Carrying out the
spin--generating transformation $\Sigma$, we arrive at the spinning
solution  
\ba\lb{spin} 
U = & \hat{\psi}'' & = p \coth \sigma - iqy \,, \nonumber \\
V = & {\ds\frac{1 + \hat{\cal E}''}{2}} & = \frac{q^2}{2} (1 -
\frac{x^2-1}{\delta^2} \sinh^2 \sigma)(1-y^2) + ipq(\frac{x}{\delta} -
\coth \sigma)y \,, \\
W = & {\ds\frac{1 - \hat{\cal E}''}{\ds 2}} & = 1 - \frac{q^2}{2} (1 -
\frac{x^2-1}{\delta^2} \sinh^2 \sigma)(1-y^2) - ipq(\frac{x}{\delta} -
\coth \sigma)y \,, \nonumber
\ea
with $p = (1+\nu^2\delta^2\Omega^2)^{-1/2}, q = \nu\delta\Omega
(1+\nu^2\delta^2\Omega^2)^{-1/2}$. Due to the presence of a non--zero $V$,
this spinning solution has a non--vanishing electromagnetic field.
However, using the asymptotic behavior $\sigma \simeq \delta/x$ for $x$
going to infinity, we check that this electromagnetic field is at most
asymptotically dipolar.

The Voorhees--Zipoy family of solutions has been generalized to a family of
spinning solutions by Tomimatsu and Sato \cite{ts} for integer $\delta$.
To compare our result (\ref{spin}) with that of Tomimatsu and Sato, we
investigate the case $\delta = 2$ (the Voorhees--Zipoy solution for
$\delta = 1$ is the \Sch solution). In this case, we obtain from
(\ref{spin}) the rescaled spinning Kinnersley potentials (to facilitate
the comparison with the Tomimatsu--Sato solution, we have multiplied the
potentials (\ref{spin}) by a common factor $2px(x^2-1)$)
\ba\lb{spin2}
U & = & p^2(x^4-1) - 2ipqx(x^2-1)y \,, \nonumber \\
V & = & - pq^2x(1-y^2) - ip^2q(x^2-1)y \,, \\
W & = & 2px(x^2-1) + pq^2x(1-y^2) + ip^2q(x^2-1)y  \nonumber
\ea
(we have checked that the corresponding Ernst potentials $\hat{\cal E}''$,
$\hat{\psi}''$ in (\ref{spin}) do indeed satisfy the Ernst equations
(\ref{ernst2})). This solution is, except for the leading terms (quartic and
cubic in $x$), obviously different from the corresponding Tomimatsu--Sato
solution \cite{ts}
\ba
U_{TS} & = & p^2(x^4-1) - 2ipqxy(x^2-y^2) - q^2(1-y^2) \,, \nonumber \\
V_{TS} & = & 0 \,, \\
W_{TS} & = & 2px(x^2-1) -2iqy(1-y^2) \,. \nonumber
\ea
From the asymptotic behaviors of the Ernst potentials of our $\delta = 2$
spinning solution, derived from (\ref{spin2}) 
\ba
{\cal E} & \simeq & 1 - \frac{4}{px} - 2iq(1+\frac{4}{p^2})\frac{y}{x^2} 
+ \cdots \,, \nonumber \\
\psi & \simeq & - iq \frac{y}{x^2} + \frac{q^2}{p} \frac{3y^2-1}{x^3}
+ \cdots 
\ea
(with $x \simeq r$, $y = \cos\theta$), we conclude that our spinning solution has, besides a monopole
gravielectric moment proportional to the mass of the original
Voorhees--Zipoy solution, a dipole gravimagnetic moment (spin), as well as
a dipole magnetic moment and a quadrupole electric moment --- rather
realistic properties for the fields generated by a configuration of
ordinary globally neutral matter.

We have given a simple procedure which generates from any asymptotically flat static axisymmetric solution of the Einstein--Maxwell equations a family of asymptotically flat spinning solutions. As an illustration, we have generated from the continuous family ($\delta$ real) of Voorhees--Zipoy solutions a family of spinning solutions differing from previously known spinning generalisations of the Voorhees--Zipoy family (e.\ g.\ the Tomimatsu--Sato solutions) in two respects: our new solutions are valid for any real $\delta$ (vs.\ integer $\delta$ for the Tomimatsu--Sato solutions); and they have non--zero electromagnetic multipole moments. In principle, similar procedures could be devised in the case of other gravitating field theories for which the stationary field equations have a high degree of symmetry, such as Kaluza--Klein theory \cite{maison2} or dilaton--axion gravity \cite{dg}.

\end{document}